\documentclass{aa}
\usepackage{epsfig,graphics}

\begin{document}

\title{Exploring the origin of magnetic fields in massive stars:\\ A survey of O-type stars in clusters  
and in the field\thanks{Based on observations obtained at the European Southern Observatory,
Paranal, Chile (ESO programme 085.D-0667(A)).}}

\author{
S.~Hubrig\inst{1}
\and
M.~Sch\"oller\inst{2}
\and
N.~V.~Kharchenko\inst{1,3}
\and
N.~Langer\inst{4}
\and
W.~J.~de Wit\inst{5}
\and
I.~Ilyin\inst{1}
\and
A.~F.~Kholtygin\inst{6}
\and
A.~E.~Piskunov\inst{1,7}
\and
N.~Przybilla\inst{8}
\and
the MAGORI collaboration
}

\institute{
Astrophysikalisches Institut Potsdam, An der Sternwarte 16, 14482~Potsdam, Germany
\and
European Southern Observatory, Karl-Schwarzschild-Str.~2, 85748~Garching, Germany
\and
Main Astronomical Observatory, 27~Academica Zabolotnogo Str., 03680~Kiev, Ukraine
\and
Argelander-Institut f\"ur Astronomie, Universit\"at Bonn, Auf dem H\"ugel~71, 53121~Bonn, Germany
\and
European Southern Observatory, Alonso de Cordova~3107, Santiago, Chile
\and
Astronomical Institute, Saint-Petersburg State University, Saint-Petersburg, Russia
\and
Institute of Astronomy of the Russian Acad.\ Sci., 48 Pyatnitskaya Str., 109017 Moscow, Russia
\and
Dr.-Karl-Remeis-Sternwarte Bamberg \& ECAP, Universit\"at Erlangen-N\"urnberg, Sternwartstr.~7, 96049~Bamberg, Germany
}

\date{Received date / Accepted date}

\abstract
{Although important effects of magnetic fields in massive stars are suggested by
recent models and observations, only a small number of massive O-type stars have been investigated 
for magnetic fields until now. Additional observations are of utmost importance to constrain the conditions 
which enable the presence of magnetic
fields and give first trends about their occurrence rate and field strength distribution.
}
{To investigate statistically whether magnetic fields in massive stars are ubiquitous
or appear in stars with specific spectral classification, certain ages, or in a special
environment, 
we acquired 41 new spectropolarimetric observations for 36 stars.
Among the observed sample roughly half of the stars are probable members of clusters at different ages, 
whereas the remaining stars are field stars not known to belong to any cluster or association.  
}
{
Spectropolarimetric observations were obtained during three different nights using the low-resolution
spectropolarimetric mode of FORS\,2 (FOcal Reducer low dispersion Spectrograph)
mounted on the 8-m Antu telescope of the VLT. To assess the membership in open clusters
and associations, we used astrometric catalogues with the best currently available kinematic and 
photometric data.
}
{
A field at a significance level of 3$\sigma$ was detected in ten O-type stars. Importantly, the largest longitudinal 
magnetic fields were measured in two Of?p stars:  $\langle$$B_z$$\rangle$\,=\,$-$381$\pm$122\,G for 
CPD$-$28\,2561 and $\langle$$B_z$$\rangle$\,=\,$-$297$\pm$62\,G for HD\,148937, previously detected by us as magnetic.
The obtained observations of HD\,148937 on three different nights
indicate that the magnetic field is slightly variable.
Our new measurements support our previous 
conclusion  that large-scale organized magnetic fields 
with polar field strengths in excess of 1\,kG are not widespread among O-type stars.
Among the stars with a detected magnetic field,
only one star, HD\,156154, belongs to an open cluster at high membership probability. According 
to previous kinematic studies, four magnetic 
O-type stars in the sample are well-known candidate runaway stars.
}
{}


\titlerunning{Magnetic fields of massive stars}
\authorrunning{S.\ Hubrig et al.}

\maketitle



\section{Introduction}
\label{sect:intro}

Magnetic fields play an important role in astrophysical phenomena of the universe at various scales.
In galaxies, dynamo models associated with various MHD instabilities occurring in the interstellar medium (ISM) 
are used  to explain the formation of the galactic structure (e.g., Gomez \& Cox \cite{GomezCox2004}; 
Bonanno \& Urpin \cite{BonannoUrpin2008}).
Magnetic fields play a role in the evolution of interstellar molecular clouds and 
in the star formation process, where
the cloud collapse is probably taking place along the magnetic field lines (e.g.\ Alves et al.\ \cite{Alves2008}).
They are also present at all stages of stellar evolution,
from young T\,Tauri stars and Ap/Bp stars
to the end products: white dwarfs, neutron stars, and magnetars.
On the other hand, role of magnetic fields
in massive O-type stars and Wolf-Rayet (WR) stars remains unknown. No definitive magnetic field was ever detected in 
WR stars and presently only less than a dozen O stars have published magnetic fields.
Also, the theories on the origin of magnetic fields in O-type stars
are still poorly developed, mostly due to the fact that the distribution of magnetic field strengths in massive stars
from the ZAMS to more evolved stages has not yet been studied.

In our study we focus on magnetic fields of massive stars observed in different environments:
in open clusters at different ages and in the field.
The results of our recent kinematic analysis of known magnetic O-type stars using the best available 
astrometric, spectroscopic, and photometric data indicates that the presence of a magnetic field 
is more frequently detected in candidate runaway stars than in stars belonging to 
clusters or associations (Hubrig et al.\ \cite{Hubrig2011b}). 
As the sample of stars with magnetic field detections 
is still very small, a study of a larger sample is urgently needed to confirm the detected trend by dedicated 
magnetic field surveys of O stars in clusters/associations and in the field. 
We were granted four nights in 2010 May with FORS\,2 at the VLT to survey magnetic fields in massive 
stars, but due to deteriorated weather conditions, only half of the granted time could be used for observations.
Notwithstanding, the obtained results allow us to preliminarily constrain the conditions which enable the 
presence of magnetic fields and give first trends about their occurrence rate
and field strength distribution. This information is critical for answering the principal question of 
the possible origin of magnetic fields in massive stars.

In the following, we present 41 new measurements of magnetic fields in 36 massive stars using FORS\,2 
at the VLT in spectropolarimetric mode.
Our observations and the obtained results are described in Sect.~\ref{sect:observations},
and their discussion is presented in Sect.~\ref{sect:discussion}.

\section{Observations and results}
\label{sect:observations}

\begin{table}
\caption{
List of O-type stars observed with FORS\,2.
Spectral classifications are listed according to the Galactic O Star Catalogue 
(Ma{\'{\i}}z-Apell{\'a}niz et al.\ \cite{maiz:2004}).}
\label{tab:objects}
\centering
\begin{tabular}{llrc}
\hline
\hline
\multicolumn{1}{c}{Name} &
\multicolumn{1}{c}{Other} &
\multicolumn{1}{c}{V} &
\multicolumn{1}{c}{Spectral Type} \\
 &
\multicolumn{1}{c}{Identifier} &
 &
 \\
\hline
CPD$-$28\,2561 & SAO\,174826     & 10.09 & O6.5 fp            \\
CPD$-$47\,2963 & SAO\,220701     & 8.45  & O4 III (f)         \\
CPD$-$58\,2620 & CD$-$58\,3529   & 9.27  & O6.5 V ((f))       \\
HDE\,303311    & CPD$-$58\,2652  & 8.98 & O5 V               \\
HD\,93129B     & CPD$-$58\,2618B & 7.16  & O3.5 V ((f+))      \\
HD\,93204      & CPD$-$59\,2584  & 8.48  & O5 V ((f))         \\
HDE\,303308     & CPD$-$59\,2623  & 8.14  & O4 V ((f+))        \\
HD\,93403      & CPD$-$58\,2680  & 7.30  & O5 III (f) var sec \\
HD\,93843      & CPD$-$59\,2732  & 7.33  & O5 III(f) var      \\
HD\,105056     & GS Mus          & 7.34  & ON9.7 Ia e         \\
HD\,115455     & CPD$-$61\,3575  & 7.97  & O7.5 III ((f))     \\
HD\,120521     & CPD$-$57\,6339  & 8.56  & O8 Ib (f)          \\
HD\,123590     & CPD$-$61\,4382  & 7.62  & O7/8               \\
HD\,125206     & CPD$-$60\,5298  & 7.94  & O9.5 IV: (n)       \\
HD\,125241     & CPD$-$60\,5300  & 8.31  & O8.5 Ib (f)        \\
HD\,130298     & CPD$-$55\,6191  & 9.29  & O6.5 III (n)(f)    \\
HD\,148937     & CPD$-$47\,7765  & 6.77  & O6.5 f?p           \\
HD\,328856     & CPD$-$46\,8218  & 8.50  & O9.5III$^*$        \\
HD\,152233     & CPD$-$41\,7718  & 6.59  & O6 III: (f)p       \\
HD\,152247     & CPD$-$41\,7732  & 7.20  & O9.5 II-III        \\
HD\,152249     & HR\,6263        & 6.47  & OC9.5 Iab          \\
HD\,153426     & CPD$-$38\,6624  & 7.49  & O9 II-III          \\
HD\,153919     & V884\,Sco       & 6.53  & O6.5 Ia f+         \\
HD\,154368     & V1074\,Sco      & 6.18 & O9.5 Iab           \\
HD\,154643     & CPD$-$34\,6733  & 7.15  & O9.5 III           \\
HD\,154811     & CPD$-$46\,8416  & 6.93  & OC9.7 Iab          \\
HD\,156154     & CPD$-$35\,6916  & 8.04  & O8 Iab (f)         \\
HD\,156212     & CPD$-$27\,5605  & 7.95  & O9.7 Iab           \\
HD\,165319     & BD$-$14\,4880   & 7.93  & O9.5Iab$^*$        \\
HD\,315033     & CPD$-$24\,6163  & 8.90  & B3$^*$             \\
HD\,168075     & BD$-$13\,4925   & 8.73  & O6 V ((f))         \\
HD\,168112     & BD$-$12\,4988   & 8.55  & O5 III (f)         \\
HD\,169582     & BD$-$09\,4729   & 8.70  & O6 I f             \\
HD\,171589     & BD$-$14\,5131   & 8.21  & O7 II (f)          \\
HD\,175754     & BD$-$19\,5242   & 7.02  & O8 II ((f))        \\
HD\,187474     & V3961\,Sgr      & 5.32 & Ap$^*$             \\
\hline
\end{tabular}
\begin{flushleft}
Notes:
$^*$ indicates Spectral Types taken from SIMBAD.
\end{flushleft}
\end{table}

Spectropolarimetric observations  were carried out
in 2010 May 20--23 in visitor mode
at the European Southern Observatory with FORS\,2 
mounted on the 8-m Antu telescope of the VLT. This multi-mode instrument is equipped with
polarization analyzing optics, comprising super-achromatic half-wave and quarter-wave 
phase retarder plates, and a Wollaston prism with a beam divergence of 22$\arcsec$  in 
standard resolution mode\footnote{
The spectropolarimetric capabilities of FORS\,1 were moved to
FORS\,2 in 2009.
}.
Polarimetric spectra were obtained with the GRISM~600B and 
the narrowest slit width of 0$\farcs$4 to achieve 
a spectral resolving power of $R\sim2000$. The use of the mosaic detector made of 
blue optimized E2V chips and a  pixel size of 15\,$\mu$m allowed us to cover a large
spectral range, from 3250 to 6215\,\AA{}, which includes all hydrogen Balmer lines 
from H$\beta$ to the Balmer jump. The spectral types and the visual magnitudes of the studied stars are 
listed in Table~\ref{tab:objects}.

A detailed description of the assessment of the longitudinal 
magnetic-field measurements using FORS\,2 is presented in our previous papers 
(e.g., Hubrig et al.\ \cite{hubrig04a,hubrig04b}, and references therein). 
The mean longitudinal 
magnetic field, $\left< B_{\rm z}\right>$, was derived using 

\begin{equation} 
\frac{V}{I} = -\frac{g_{\rm eff} e \lambda^2}{4\pi{}m_ec^2}\ \frac{1}{I}\ 
\frac{{\rm d}I}{{\rm d}\lambda} \left<B_{\rm z}\right>, 
\label{eqn:one} 
\end{equation} 

\noindent 
where $V$ is the Stokes parameter that measures the circular polarisation, $I$ 
is the intensity in the unpolarised spectrum, $g_{\rm eff}$ is the effective 
Land\'e factor, $e$ is the electron charge, $\lambda$ is the wavelength, $m_e$ the 
electron mass, $c$ the speed of light, ${{\rm d}I/{\rm d}\lambda}$ is the 
derivative of Stokes $I$, and $\left<B_{\rm z}\right>$ is the mean longitudinal magnetic 
field. 

Longitudinal magnetic fields were measured in two ways: using only the absorption hydrogen Balmer 
lines or using the entire spectrum including all available absorption lines.
The lines that show evidence for emission were not used in the 
determination of the magnetic field strength. 
The feasibility of longitudinal magnetic field measurements in massive stars 
using low-resolution spectropolarimetric observations was demonstrated by previous studies of O and B-type stars
(e.g., Hubrig et al.\ \cite{hubrig:2006,hubrig08,hubrig09,Hubrig2011a}).
To check that the instrument was functioning properly, we observed the magnetic Ap star HD\,187474, which has a 
well studied longitudinal magnetic field, during the night of May~23 at rotation phase 0.66.
HD\,187474 has a rotation period of 6.4\,yr and a longitudinal magnetic field ranging roughly from 
$-$2\,kG to 2\,kG. The measured value 
of the magnetic field, $\left< B_z\right>_{\rm all}= -1249\pm47$\,G, fits very well to the observations 
at the same phase presented by Landstreet \& Mathys (\cite{LandstreetMathys2000}).

\begin{table*}
\caption{
Longitudinal magnetic fields measured with FORS\,2 in the studied sample.
All quoted errors are 1$\sigma$ uncertainties.
}
\label{tab:fields}
\centering
\begin{tabular}{lcr @{$\pm$} lr @{$\pm$} lc}
\hline
\hline
\multicolumn{1}{c}{Name} &
\multicolumn{1}{c}{MJD} &
\multicolumn{2}{c}{$\left< B_z\right>_{\rm all}$} &
\multicolumn{2}{c}{$\left< B_z\right>_{\rm hydr}$} &
\multicolumn{1}{c}{Comment} \\
 &
 &
\multicolumn{2}{c}{[G]} &
\multicolumn{2}{c}{[G]} &
 \\
\hline
CPD$-$28\,2561 & 55338.969 &  $-$381 & 122 &  $-$534 & 167 & ND \\
CPD$-$47\,2963 & 55337.094 &  $-$190 &  62 &  $-$154 & 96 & ND \\
CPD$-$58\,2620 & 55339.020 &   $-$53 &  71 &   $-$66 &  88 & \\
HDE\,303311    & 55337.131 &   $-$56 &  40 &   $-$19 &  61 & \\
HD\,93129B     & 55337.179 &   $-$49 &  44 &   $-$79 &  78 & \\
HD\,93204      & 55339.053 &      22 &  46 &      16 &  66 & \\
HDE\,303308     & 55339.078 &     122 &  54 &     137 &  96 & \\
HD\,93403      & 55339.116 &      39 &  41 &   $-$15 &  88 & \\
HD\,93843      & 55339.099 &  $-$157 &  42 &  $-$173 &  56 & ND \\
HD\,105056     & 55337.208 &   $-$93 &  48 &  $-$156 &  63 &  \\
HD\,115455     & 55337.230 &       4 &  47 &      13 &  64 & \\
HD\,120521     & 55337.257 &      25 &  44 &    $-$3 &  62 & \\
HD\,123590     & 55339.133 &      19 &  42 &      56 &  70 & \\
HD\,125206     & 55337.307 &      12 &  64 &       8 &  91 & \\ 
HD\,125241     & 55339.188 &      24 &  39 &      16 &  72 & \\
HD\,130298     & 55339.159 &     113 &  38 &     193 &  62 & ND \\
HD\,148937     & 55336.307 &  $-$297 &  62 &  $-$293 &  85 & CD \\
               & 55337.285 &  $-$204 &  71 &  $-$225 & 103 & \\
               & 55339.206 &  $-$290 &  85 &  $-$389 & 129 & \\
HD\,328856     & 55336.370 &  $-$173 &  53 &  $-$155 &  65 & ND \\
               & 55339.223 &  $-$149 &  48 &   $-$75 &  72 & \\
HD\,152233     & 55339.289 &   $-$74 &  52 &   $-$76 &  74 & \\
HD\,152247     & 55339.261 &      34 &  52 &      86 &  64 & \\
HD\,152249     & 55339.275 &   $-$22 &  51 &   $-$15 &  72 & \\
HD\,153426     & 55336.338 &   $-$27 &  53 &   $-$10 &  62 & \\
               & 55339.246 &  $-$171 &  55 &  $-$275 &  70 & ND \\
HD\,153919     & 55337.341 &  $-$213 &  68 &  $-$119 &  95 & ND\\
HD\,154368     & 55339.324 &   $-$74 &  38 &   $-$77 &  63 & \\
HD\,154643     & 55339.340 &     110 &  34 &     121 &  52 & ND \\
HD\,154811     & 55337.327 &      91 &  39 &      59 &  64 & \\
HD\,156154     & 55337.358 &  $-$118 &  38 &  $-$167 &  54 & ND \\
HD\,156212     & 55337.377 &  $-$104 &  42 &   $-$51 &  63 & \\
HD\,165319     & 55339.306 &   $-$44 &  48 &   $-$38 &  74 & \\ 
HD\,315033     & 55337.402 &   $-$41 &  41 &   $-$36 &  52 & \\
HD\,168075     & 55339.389 &      17 &  42 &       3 &  65 & \\
HD\,168112     & 55336.418 &   $-$74 &  53 &   $-$66 &  74 & \\
               & 55339.362 &     112 &  62 &      40 &  96 & \\
HD\,169582     & 55339.415 &   $-$87 &  58 &  $-$124 &  88 & \\
HD\,171589     & 55339.445 &   $-$62 & 140 &  $-$119 & 180 & \\
HD\,175754     & 55337.426 &      82 &  67 &      66 &  77 & \\
\hline
HD\,187474     & 55339.433 & $-$1249 &  22 & $-$1253 &  32 & Ap star \\
\hline
\end{tabular}
\end{table*}

Although we were granted four nights for our survey, due to unfavorable weather conditions 
(snow and high humidity) only four stars could be observed during the first night, 14 stars 
during the second night, none during the third night, and 23 during the last night. 
Most of the targets were observed only once. The 
exceptions were the stars HD\,328856, HD\,153426, and HD\,168112, which we were able to observe twice.
HD\,148937 was observed three times to assess the magnetic field variability over the rotation cycle.
Apart from this star, which has a rotation period of seven days (Naz\'e et al.\ \cite{naze08}), 
no exact rotation periods are known for the other stars in our sample. 

The results of our magnetic field measurements are presented in Table~\ref{tab:fields}.
In the first two columns, we provide the star names and the modified Julian dates at the middle of 
the exposures. 
In Cols.~3 and 4 we present the longitudinal magnetic 
field $\left<B_{\rm z}\right>_{\rm all}$ using the whole spectrum and the longitudinal magnetic field 
$\left<B_{\rm z}\right>_{\rm hyd}$ using only the hydrogen lines. 
All quoted errors are 1$\sigma$ uncertainties.
In Col.~5, we identify new detections by ND and in the case of HD\,148937 the confirmed detection
is marked by CD.  

Ten stars of our sample, CPD$-$28\,2561, CPD$-$47\,2963, HD\,93843, HD\,130298, HD\,148937, HD\,328856, 
HD\,153426, HD\,153919, HD\,154643, and HD\,156154, 
show evidence for the presence of a magnetic field. 

Importantly, the strongest magnetic fields are detected in both Of?p stars CPD$-$28\,2561 and  HD\,148937.
Walborn (\cite{walb73})
introduced the Of?p category for massive O stars displaying recurrent spectral variations in 
certain spectral lines, sharp emission or P Cygni profiles in He~I and the Balmer lines, and strong C~III 
emission lines around 4650\,\AA{}.
Only five Galactic Of?p stars are presently known: HD\,108, NGC\,1624-2, CPD$-$28\,2561, HD\,148937, and HD\,191612 
(Walborn et al.\ \cite{walb10}). 
Our observations of CPD$-$28\,2561 reveal a magnetic field at 3.1$\sigma$ level using the whole spectrum and 
at 3.2$\sigma$ level using Balmer lines.
The study  of radial velocity variation of Levato et al.\ (\cite{Levato1988}) indicated the presence of variability 
of a few emission lines with a probable period of 17 days.  
Walborn et al.\ (\cite{walb10}) report that CPD$-$28\,2561 undergoes extreme spectral transformations very similar 
to those of HD\,191612, on a timescale of weeks, exhibiting variable emission intensity 
of the C~III $\lambda\lambda$~4647-4650-4652 triplet. 
The detection of a mean longitudinal magnetic field $\langle$$B_z$$\rangle$\,=\,$-$254$\pm$81\,G
in the Of?p star HD\,148937 using FORS\,1 at the VLT was previously reported by Hubrig et al.\ (\cite{hubrig08}).
An extensive multiwavelength study of HD\,148937 was carried out by Naz\'e et al.\ (\cite{naze08}), who
detected small-scale variations of \ion{He}{ii} 4686 and the Balmer lines with a period of 7\,days and an overabundance 
of nitrogen by a factor of 4 compared to the Sun. The periodicity of spectral variations in hydrogen and He lines
was re-confirmed using additional higher resolution spectroscopic material indicating similarity to the other Of?p stars
HD\,108 and HD\,191612 (Naze et al.\ \cite{Naze2010}).
Our spectropolarimetric observations of this star indicate that the magnetic field is variable, but due to the low number of 
measurements it is not possible to verify the period deduced from spectroscopic observations. 
The magnetic field of this star was observed at 4.8$\sigma$, 2.9$\sigma$, 
and 3.4$\sigma$ levels on three different nights using all absorption lines. 

The remaining three Of?p stars are located in the Northern hemisphere and cannot be reached with FORS\,2 at the VLT.
To study magnetic fields in HD\,108 and HD\,191612 we recently used  polarimetric spectra obtained with the SOFIN 
spectrograph installed at the Nordic Optical Telescope (Hubrig et al.\ \cite{Hubrig2010}).
As a result, we detected a longitudinal magnetic field $\langle$$B_z$$\rangle$\,=\,$-$168$\pm$35\,G in the Of?p star HD\,108, 
which is in agreement with the longitudinal magnetic field measurement of 
the order of $-$150\,G recently reported by Martins et al.\ (\cite{Martins2010}).
For the star HD\,191612 with a rotation period of 537.6\,d (Howarth et al.\ \cite{Howarth2007}) we measured a longitudinal magnetic field 
$\langle$$B_z$$\rangle$\,=\,450$\pm$153\,G at rotation phase 0.43 (Hubrig et al.\ \cite{Hubrig2010}).
The only previously 
published magnetic field measurement for this star showed a negative longitudinal magnetic 
field $\langle$$B_z$$\rangle$\,=\,$-$220$\pm$38\,G
at rotation phase 0.24 (Donati et al.\ \cite{donati:2006}), indicating a change of 
polarity over $\sim$100 days. 
No attempt has yet been made to measure the magnetic field of NGC\,1624-2.
Clearly the recent results of magnetic field measurements in Of?p stars imply a tight relation between the 
observed properties of the Of?p star group and the presence of a magnetic field. 

For the star CPD$-$47\,2963 we achieved a
3.1$\sigma$ detection using all absorption lines.
According to Walborn et al.\ (\cite{walb10}) this star belongs to the Ofc category, which consists of normal spectra 
with C~III $\lambda\lambda$~4647-4650-4652 emission lines of comparable intensity to those of the Of defining lines 
N~III $\lambda\lambda$~4634-4640-4642. The authors indicate that the Ofc phenomenon occurs primarily in 
certain associations and young clusters. However, the available kinematic and photometric data do not indicate 
cluster or association membership for CPD$-$47\,2963. 
The origin of the magnetic field in this star is probably different compared to that of other magnetic
O-type stars, as non-thermal radio emission, which is frequently observed
in binary systems with colliding winds, was detected by Benaglia et al.\ (\cite{Benaglia2001}). 
On the other hand, the membership of
CPD$-$47\,2963 in a binary or multiple system has not been investigated yet. The authors suggest that the 
non-thermal radiation from this star possibly comes from strong shocks in the wind itself and/or in the 
wind colliding region if the star has a massive early-type companion. Both optical and radio observations
reveal the presence of a second source separated by 5\arcsec{}.

According to Walborn et al.\ (\cite{walb10}), also the star HD\,93843, with a 3.7$\sigma$ detection 
achieved using all absorption lines, belongs to the Ofc category.
Prinja et al.\ (\cite{Prinja1998}) monitored the stellar wind of this star using IUE time series.
They identified systematic changes in the absorption troughs of the \ion{Si}{iv} and \ion{N}{v} resonance lines with a repeatability 
of wind structures with a period of 7.1\,days. Noteworthy, the authors suggest the presence
of a magnetic field as one of the possible mechanisms to explain the cyclical wind perturbation.
On the other hand, three other stars of the  Ofc category included in our survey,
HD\,93204, HDE\,303308, and HD\,93403, do not show the presence of a magnetic field at a 3$\sigma$ level.

The star HD\,130298 with a longitudinal magnetic field observed at 3.1$\sigma$ level using the Balmer lines,
is known as an object with a bow shock. Noriega-Crespo et al.\ (\cite{NoriegaCrespo1997}) used ISSA/IRAS 
archival spectra  to identify stars surrounded
by extended infrared emission at 60$\mu$m, which is a signature of wind bow shocks.
The bow shocks are usually associated with runaway early-type stars with 
typical wind velocities of 500--3000\,km\,s$^{-1}$ and mass loss rates $\sim10^{-5}$ -- $10^{-6}$ $M_{\sun}$\,yr$^{-1}$
(see e.g.\ Puls et al.\ \cite{Puls1996}).

\begin{table*}
\caption{
Probable members in open clusters and associations. 
}
\label{tab:clusters}
\centering
\begin{tabular}{lrrrlrrr@{,\,}r@{,\,}rr@{,\,}r@{,\,}r}
\hline
\hline
\multicolumn{1}{c}{Object} &
\multicolumn{1}{c}{ASCC} &
\multicolumn{1}{c}{$P_{\rm kin}$} &
\multicolumn{1}{c}{$P_{\rm phot}$} &
\multicolumn{1}{c}{Cluster} &
\multicolumn{1}{c}{dist} &
\multicolumn{1}{c}{log\,$t$}& 
\multicolumn{3}{c}{$(\mu_X$, $\mu_Y$, $\sigma_{\mu})^{\rm star}$} &
\multicolumn{3}{c}{$(\mu_X$, $\mu_Y$, $\sigma_{\mu})^{\rm cl}$}\\
\multicolumn{1}{c}{name} &
\multicolumn{1}{c}{number} &
\multicolumn{1}{c}{[\%]} &
\multicolumn{1}{c}{[\%]} &
\multicolumn{1}{c}{ } &
\multicolumn{1}{c}{[pc]} &
\multicolumn{1}{c}{[yr]} &
\multicolumn{3}{c}{[mas/yr]} &
\multicolumn{3}{c}{[mas/yr]} \\
\hline
 HD\,47839$^{\ast}$       & 1021435 & 62 & 100 & NGC\,2264    &  660 & 6.81       & $-$3.84 & $-$2.50 & 0.94 &  $-$2.70 &  $-$3.50 & 0.25 \\
 CPD$-$58\,2620           & 2232461 & 44 &   1 & Trumpler\,14 & 2753 & 6.67       & $-$3.77 & $-$2.20 & 2.37 &  $-$3.91 &     3.65 & 0.60 \\
 HDE\,303311              & 2232607 & 91 &  27 & Trumpler\,14 & 2753 & 6.67       & $-$2.91 &    2.64 & 2.40 &  $-$3.91 &     3.65 & 0.60 \\
 HD\,93129B               & 2232449 &  0 & 100 & Trumpler\,14 & 2753 & 6.67       & $-$9.80 &   10.86 & 1.96 &  $-$3.91 &     3.65 & 0.60 \\
 HD\,93204                & 2232588 & 75 &   0 & Trumpler\,16 & 2842 & 6.90       & $-$9.33 &    3.16 & 1.70 & $-$11.10 &     4.02 & 0.42 \\
 HDE\,303308              & 2232708 & 19 &   0 & Trumpler\,16 & 2842 & 6.90       & $-$6.44 &    2.54 & 1.79 & $-$11.10 &     4.02 & 0.42 \\
 HD\,105056               & 2396832 & 12 & 100 & ASCC\,69     & 1000 & 7.91       & $-$4.56 & $-$1.81 & 0.87 &     7.91 &  $-$7.52 & 0.41 \\
 HD\,115455               & 2338543 & 90 &  72 & Stock\,16    & 1640 & 6.78       & $-$2.56 &    1.08 & 2.02 &  $-$3.31 &  $-$0.03 & 0.66 \\
 HD\,120521               & 2256830 &  0 & 100 & Platais\,10  &  246 & 8.20       & $-$5.55 &    0.40 & 1.39 & $-$29.10 & $-$10.73 & 0.40 \\
 HD\,123590               & 2345821 & 83 &  83 & ASCC\,77     & 2200 & 6.99       & $-$4.96 & $-$2.45 & 0.99 &  $-$4.59 &  $-$1.49 & 0.47 \\
 HD\,328856               & 2074567 & 22 & 100 & Hogg\,22     & 1297 & 6.70       &    0.78 & $-$1.37 & 1.26 &  $-$0.84 &  $-$4.39 & 0.43 \\
 HD\,135240$^{\ast}$      & 2351294 & 61 & 100 & ASCC\,79     &  800 & 6.86       & $-$3.00 & $-$2.65 & 0.85 &  $-$2.67 &  $-$4.10 & 0.44 \\
 HD\,135591$^{\ast}$      & 2351425 & 67 & 100 & ASCC\,79     &  800 & 6.86       & $-$3.00 & $-$2.65 & 0.85 &  $-$2.67 &  $-$4.10 & 0.44 \\
 HD\,152233               & 1974789 & 13 & 100 & NGC\,6231    & 1250 & 6.81       & $-$4.11 & $-$8.05 & 2.40 &  $-$0.39 &  $-$1.99 & 0.46 \\
 HD\,152247               & 1974811 & 95 & 100 & NGC\,6231    & 1250 & 6.81       & $-$0.55 & $-$1.14 & 1.91 &  $-$0.39 &  $-$1.99 & 0.46 \\
 HD\,152249               & 1974812 & 84 & 100 & NGC\,6231    & 1250 & 6.81       &    0.13 & $-$0.81 & 1.36 &  $-$0.39 &  $-$1.99 & 0.46 \\
 HD\,153919               & 1877055 &  0 & 100 & NGC\,6281    &  494 & 8.51       &    1.42 &    4.84 & 1.04 &  $-$2.96 &  $-$3.75 & 0.23 \\
 HD\,154368               & 1877523 & 88 & 100 & ASCC\,88     & 1900 & 7.17       &    3.80 & $-$2.08 & 1.05 &     2.89 &  $-$2.00 & 0.32 \\
 {\it HD\,155806}$^{\ast}$& 1783567 & 98 & 100 & Sco\,OB4     & 1100 & 6.82       &    0.33 & $-$2.02 & 0.80 &     0.46 &  $-$2.22 & 0.15 \\
 {\it HD\,156154}         & 1878846 & 82 & 100 & Bochum\,13   & 1077 & 7.08       & $-$0.88 & $-$2.53 & 1.64 &  $-$0.28 &  $-$1.20 & 0.30 \\
 {\it HD\,164794}$^{\ast}$& 1593528 & 69 & 100 & NGC\,6530    & 1322 & 6.67       &    1.92 & $-$0.40 & 1.09 &     2.01 &  $-$1.81 & 0.51 \\
 HD\,315033               & 1593655 & 97 & 100 & NGC\,6530    & 1322 & 6.67       &    2.33 & $-$2.51 & 2.09 &     2.01 &  $-$1.81 & 0.51 \\
 HD\,167263$^{\ast}$      & 1595683 & 97 & 100 & Sgr\,OB7     & 1860 & 6.64       &    1.56 & $-$1.53 & 1.19 &     1.80 &  $-$1.20 & 0.26 \\
 HD\,168075               & 1407400 & 73 & 100 & NGC\,6611    & 1719 & 6.72       &    2.81 & $-$1.14 & 1.18 &     1.60 &  $-$0.35 & 0.41 \\
 HD\,168112               & 1407419 & 73 &  19 & NGC\,6604    & 1696 & 6.64       &    1.67 & $-$2.20 & 1.69 &  $-$0.36 &  $-$2.68 & 0.31 \\
 \hline
 HD\,153426               & 1876676 &  0 &   0 & Hogg\,22     & 1297 & 6.70(6.91) & $-$0.21 & $-$0.09 & 1.38 &  $-$0.84 &  $-$4.39 & 0.43 \\
 HD\,154643               & 1877779 &  0 &   0 & ASCC\,88     & 1900 & 7.17(6.15) &    3.66 & $-$1.71 & 1.55 &     2.89 &  $-$2.00 & 0.32 \\
 HD\,169582               & 1323022 &  0 &   0 & NGC\,6604    & 1696 & 6.64(6.74) & $-$0.40 & $-$1.63 & 1.14 &  $-$0.36 &  $-$2.68 & 0.31 \\
 HD\,171589               & 1409170 &  0 &   0 & NGC\,6618    & 1814 & 7.78(6.52) &    5.10 &    3.74 & 1.66 &     3.04 &     0.33 & 0.92 \\
 HD\,175754               & 1503597 &  0 &   0 & ASCC\,93     & 2500 & 7.72(7.01) &    1.49 &    2.15 & 1.01 &  $-$2.18 &  $-$1.80 & 0.55 \\
\hline
\end{tabular}
\tablefoot{
 For each star we give in the first two columns the object name and the corresponding catalogue number.
The kinematic and photometric probabilities for cluster membership presented in Cols.~3 und 4 were calculated 
according to the procedures described by Kharchenko et al.\ (\cite{Kharchenko2004}). 
The cluster names, the distances and
the ages are presented in the next three columns. The proper motions and their errors for stars and clusters 
are given in
the last columns. All cluster data are taken from Kharchenko et al.\ (\cite{Kharchenko2005a, Kharchenko2005b}).
The five stars at the end of the table are currently not members of the indicated clusters,
and were ejected from them $t_{\rm enc}$ years ago 
(Schilbach,  Roeser~\cite{SchilbachRoeser2008}). The encounter times $t_{\rm enc}$ of these stars are given in brackets.
The three stars with magnetic fields that are most probable cluster members are presented in italics.
}
\end{table*}

The two stars HD\,328856 and HD\,153426, both with magnetic field detections, were observed on two different nights,
namely the first and the fourth night of our observing run.
For HD\,328856 we obtained on these nights 3.3$\sigma$ and 3.1$\sigma$ level
detections, respectively, using all absorption lines.
Based on the photometric membership probability, this star is a member of the compact open
cluster Hogg\,22 in the Ara region at an age of 5\,Myr and a
distance of about 1300\,pc (see Sect. 3). On the other hand, its proper motions
indicate that HD\,328856 is not fully co-moving with the other cluster members, deviating from the cluster
mean proper motion by $\sim$2$\sigma$ (for more details on membership probabilities
see Kharchenko et al.\ \cite{Kharchenko2004}).
The observations of the star HD\,153426 revealed the presence of a mean longitudinal magnetic field
at the 3.9\,$\sigma$ level using Balmer lines on the fourth night. The non-detection
of the magnetic field for HD\,153426 on the first observing night can probably be explained by the
strong dependence of the longitudinal magnetic field on the rotational aspect.
HD\,153426 is a double-lined spectroscopic binary with unknown orbit parameters and was considered by
de Wit et al.\ (\cite{deWit2005}) as a star in a newly detected cluster. Using the best presently available
kinematic data on young open clusters, Schilbach \& R\"oser (\cite{SchilbachRoeser2008})
suggested that HD\,153426 was ejected from the cluster Hogg\,22.
Their back-tracing procedure indicates that the
encounter time for HD\,153426, i.e.\ the time when the star was ejected, is about 8.1\,Myr,
while the age of the cluster Hogg\,22 is only 5\,Myr.

The star HD\,153919 was observed only once, revealing the presence of a mean longitudinal magnetic field at the
3.9\,$\sigma$ level, using all absorption lines. The study of Ankay et al.\ (\cite{Ankay2001}) suggested that
this star is a runaway X-ray binary,
ejected from the OB association Sco OB1 about 2\,Myr ago due to the supernova of 4U1700-37's progenitor.
They considered this star as a companion to 4U1700-37, most likely a
neutron star powered by wind accretion (e.g., Jones et al.\ \cite{Jones1973}). Since 4U1700-37 is a
candidate for a low-mass black hole (Brown et al.\ \cite{Brown1996}), this system can be similar to
the optical component (the O9.7 Iab supergiant) in the system Cyg X-1, for which the presence of a variable weak
magnetic field was recently detected using a FORS\,1 spectropolarimetric time series over the orbital period of 5.6\,days
(Karitskaya et al.\ \cite{Karitskaya2010}). Schilbach \& R\"oser (\cite{SchilbachRoeser2008})
identified the origin of this field star in the cluster NGC\,6231 (the open cluster inside Sco~OB1) at
an age of about 6.5\,Myr and their back-tracing procedure indicates that the
star was ejected from the cluster 1.1\,Myr ago.

The longitudinal magnetic field for the star HD\,154643 was observed at 3.2$\sigma$ level using all absorption lines.
De Wit et al.\ (\cite{deWit2005}) characterise this star as a
candidate runaway star associated with the young cluster Bochum\,13.
However, Schilbach \& R\"oser (\cite{SchilbachRoeser2008})
identified the origin of this field star in the cluster ASCC\,88 at an age of about 14.8\,Myr, ejected
1.4\,Myr ago.

The star HD\,156154, for which we achieved a 3.1$\sigma$ detection using all absorption lines, seems
to be the only star with a high
cluster membership probability among the presented O-type stars with detected magnetic fields (see Sect.~3).
According to kinematic and photometric criteria belongs this star to the open cluster Bochum\,13
at an age of 12\,Myr and a distance of about 1\,kpc.

\section{Discussion}
\label{sect:discussion}

A lot of effort has been put into the research of massive stars in
recent years in order to properly model the effects of rotation, stellar
winds, and surface chemical composition. However, possible paths for the formation of
magnetic O-type stars were not analysed yet with modern theories for the evolution of single and binary stars.
Clearly, the number of massive stars with detected magnetic fields is still small, and the available data
are insufficient to prove statistically whether magnetic fields in massive stars are ubiquitous
or appear in specific stars with certain stellar parameters and in a special
environment. On the other hand, the observations of magnetic fields in massive stars accumulated over the last few years
can be used to preliminarily constrain the conditions which enable the appearance of magnetic fields and give first
trends about their occurrence and field strength distribution.

Since no longitudinal magnetic fields stronger than 300\,G were detected in our study
(apart from the rather large field in the Of?p star CPD$-$28\,2561), we confirm our previous conclusion
(Hubrig et al.\ \cite{hubrig08}) that
large-scale, dipole-like, magnetic fields with polar field strengths
higher than 1\,kG are not widespread among O type stars.
Our study presents the results of a magnetic field survey in 36 massive stars. Among them, 
19 stars can be related to open clusters and associations at different age. The data on the cluster 
membership of
these probable cluster O-type stars are presented in Table~\ref{tab:clusters}.
To increase the significance of our statistic assessment, we present in the same table the data for an
additional six probable cluster O-type stars, which have been studied during
the last years by Hubrig et al.\ (\cite{hubrig08,Hubrig2009,Hubrig2011b}),
marked for convenience by an asterisk.
As database for the compilation of Table~\ref{tab:clusters} we used the All-sky Compiled Catalogue of 2.5 million 
stars (ASCC-2.5, 3rd version) of Kharchenko \& Roeser (\cite{KharchenkoRoeser2009}).
We note that for the calculation of kinematic membership probability only proper motions were used. 
According to Dias et al.\ (\cite{Dias2002}; Version 3.1 (24/11/2010)),
one of the previously studied O-type stars, the star HD\,152408 (Hubrig et al.\ \cite{hubrig08}),
is projected on the cluster Collinder\,316, which presents a large group of bright stars superposed on 
Trumpler\,24 at the age log\,$t$= 6.92. We have not included this star in Table~\ref{tab:clusters}, as no membership 
criteria are discussed in this work. 
According to Mason et al.\ (\cite{Mason1998}) and
Pourbaix et al.\ (\cite{Pourbaix2004})\footnote{http//sb9.astro.ulb.ac.be/mainform.cgi}, 
among the stars presented in Table~\ref{tab:clusters}, six stars, HD\,47839, HD\,135240, HD\,152233, HD\,153919, HD\,154368, and
HD\,167263, are members of spectroscopic binary systems, with orbital periods between 3.4 and 9247 days.  

\begin{figure}
\centering
\includegraphics[height=0.45\textwidth,angle=270]{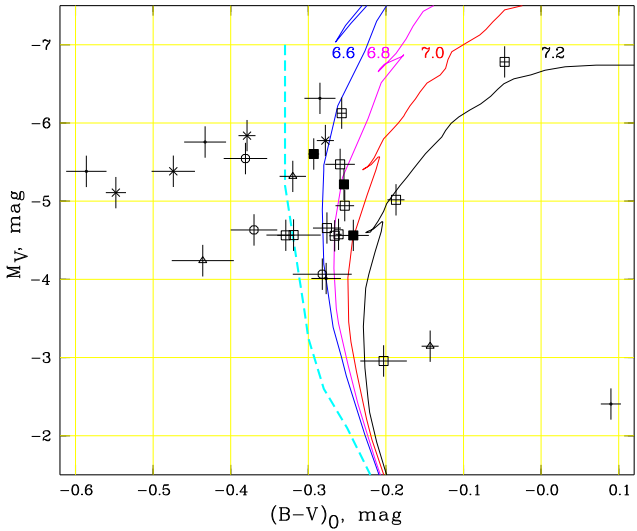}
\caption{
The positions of the stars studied for cluster membership in the colour-magnitude diagram. Different symbols indicate
stars with different membership probabilities:
Squares stand for stars with cluster membership probability larger than 60\%,
circles for stars with a probability between 14\% and 60\%,
and triangles for membership probability between 1\% and 14\%.
Non-members and runaway stars are marked by dots and crosses, respectively.
One runaway star, HD\,171589, does not appear in this figure, as its colour and
magnitude do not fit the presented parameter space (see Table~\ref{tab:colours}).
The three stars with magnetic fields, HD\,155806, HD\,156154, and HD\,164794, with high cluster membership probabilities
are denoted by filled squares. Isochrones for log\,$t=$6.6, 6.8, 7.0, and 7.2 are presented by solid lines and the zero-age 
main sequence is shown as a dashed line.
}
\label{fig:evol}
\end{figure}

\begin{figure}
\centering
\includegraphics[width=0.45\textwidth]{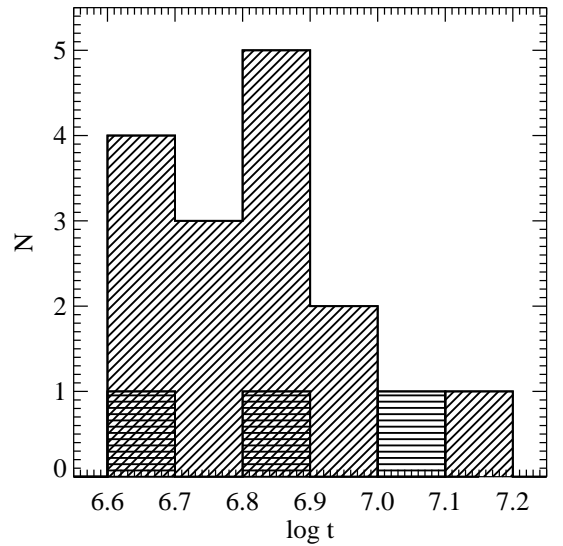}
\caption{
The age distribution of studied probable cluster members.
The three stars with magnetic fields, HD\,155806, HD\,156154, and HD\,164794, are denoted by horizontal lines.
}
\label{fig:histo}
\end{figure}

From the inspection of kinematic and
photometric membership probabilities, and following the membership criteria described by Kharchenko et al.\ (\cite{Kharchenko2004}),
five stars in Table~\ref{tab:clusters},  CPD$-$58\,2620, HD\,93129B, HDE\,303308, HD\,105056,
HD\,152233, have a rather low cluster membership probability. Two other stars, HD\,120521 and HD\,153919, 
are not kinematic members of the oldest clusters Platais 10 and NGC 6281, respectively, and should be regarded as 
field stars projected against the clusters by chance.
Among the remaining stars, only in three stars, HD\,155806, HD\,156154, and HD\,164794, weak magnetic fields
have been detected (more details on the kinematic study of HD\,155806 and HD\,164794 can be found in 
Hubrig et al.\ \cite{Hubrig2011b}).
In Fig.~\ref{fig:evol} we display the positions of the stars listed in Table~\ref{tab:clusters} in the colour-magnitude 
diagram. The absolute magnitudes $M_{\rm V}$ and intrinsic colours $(B-V)_0$ were calculated using cluster distances 
and $E(B-V)$ values presented by Kharchenko et al.\ (\cite{Kharchenko2005a, Kharchenko2005b}). 
The theoretical isochrones were calculated by the Padova group 
(Girardi et al.\ \cite{Girardi2002}) and the values for the zero-age main sequence were 
retrieved from Schmidt-Kaler (\cite{SchmidtKaler1982}).
The absolute magnitudes and colours are presented in Table~\ref{tab:colours}.
The uncertainty in the distance modulus is assumed to be 0.2.
The positions of the magnetic stars with high cluster membership 
probabilities, HD\,156154, HD\,155806, and HD\,164794, do not reveal any sign of a specific distribution, which
would hint at the origin of their magnetic fields at a certain evolutionary age. 

\begin{table}
\caption{
Absolute magnitudes and intrinsic colours of stars studied for the cluster membership. 
}
\label{tab:colours}
\centering
\begin{tabular}{lrr@{$\pm$}l}
\hline
\hline
\multicolumn{1}{c}{Object name} &
\multicolumn{1}{c}{$M_{\rm V}$} &
\multicolumn{2}{c}{$(B-V)_0$} \\
\hline
 HD\,47839      & $-$4.557 & $-$0.266 & 0.007 \\
 CPD$-$58\,2620 & $-$4.239 & $-$0.436 & 0.040 \\
 HDE\,303311    & $-$4.632 & $-$0.370 & 0.030 \\
 HD\,93129B     & $-$6.315 & $-$0.285 & 0.020 \\
 HD\,93204      & $-$5.381 & $-$0.586 & 0.026 \\
 HDE\,303308    & $-$5.756 & $-$0.433 & 0.027 \\
 HD\,105056     & $-$3.145 & $-$0.143 & 0.011 \\
 HD\,115455     & $-$4.563 & $-$0.329 & 0.011 \\
 HD\,120521     & $-$4.011 & $-$0.277 & 0.019 \\
 HD\,123590     & $-$5.017 & $-$0.187 & 0.011 \\
 HD\,328856     & $-$4.066 & $-$0.282 & 0.038 \\
 HD\,135240     & $-$4.942 & $-$0.253 & 0.012 \\
 HD\,135591     & $-$4.576 & $-$0.261 & 0.004 \\
 HD\,152233     & $-$5.317 & $-$0.320 & 0.017 \\
 HD\,152247     & $-$4.656 & $-$0.276 & 0.018 \\
 HD\,152249     & $-$5.471 & $-$0.259 & 0.019 \\
 HD\,153919     & $-$2.406 &    0.090 & 0.013 \\
 HD\,154368     & $-$6.781 & $-$0.047 & 0.006 \\
 HD\,155806     & $-$5.214 & $-$0.254 & 0.006 \\
 HD\,156154     & $-$4.562 & $-$0.242 & 0.020 \\
 HD\,164794     & $-$5.603 & $-$0.293 & 0.007 \\
 HD\,315033     & $-$2.956 & $-$0.203 & 0.030 \\
 HD\,167263     & $-$6.125 & $-$0.257 & 0.006 \\
 HD\,168075     & $-$4.566 & $-$0.319 & 0.035 \\
 HD\,168112     & $-$5.545 & $-$0.381 & 0.028 \\
 \hline
 HD\,153426     & $-$5.109 & $-$0.548 & 0.013 \\
 HD\,154643     & $-$5.775 & $-$0.278 & 0.011 \\
 HD\,169582     & $-$5.382 & $-$0.474 & 0.028 \\
 HD\,171589     & $-$8.036 & $-$1.340 & 0.021 \\
 HD\,175754     & $-$5.838 & $-$0.379 & 0.011 \\
\hline
\end{tabular}
\end{table}

In Fig.~\ref{fig:histo} we present the age distribution of the most probable cluster members. While the age of 
HD\,155806 and HD\,164794 is similar to the bulk of the studied cluster O-type stars, the star HD\,156154 is somewhat older
at an age of $\sim$12\,Myr.
HD\,155806 is classified as an Oe star, possibly representing 
the higher mass analogues of classical Be stars (e.g.\ Walborn \cite{walb73}). Only six members
are suggested to belong to this group of stars (e.g.\ Negueruela et al.\ \cite{neg04}).
The star HD\,164794 is a spectroscopic double-lined system with an orbital period of 2.4\,yr, known as
emitting non-thermal radio-emission, probably associated with colliding winds (Naz\'e et al.\ \cite{Naze2010}).
No specific information can be found in the literature about the luminous supergiant HD\,156154.

The available observations seem to indicate that the presence of a magnetic field 
is more frequently detected in field stars than in stars belonging to 
clusters or associations. It is generally accepted that the majority of massive stars form in star clusters and associations, and 
studies of kinematical properties of the massive star field population indicate that a major
part 
of these stars can be traced back to their parent open clusters or associations (e.g.\ Schilbach \& Roeser 2008).
Pflamm-Altenburg \& Kroupa (\cite{PflammAltenburgKroupa2010})
recently discussed in their work whether massive stars can form in isolation in the galactic field. According to 
de Wit et al.\ (\cite{deWit2005}), only a few per cent of all O-type stars can be considered as formed outside a 
cluster environment. Pflamm-Altenburg \& Kroupa considered the two-step-ejection process, which presents the combination 
of the dynamical and the supernova ejection scenario with the result that massive field stars produced via 
this ejection process for the vast majority of cases cannot be traced back to their parent star clusters. 
These stars can be mistakenly considered as massive stars formed in isolation. While this can not be proven, the 
observed numbers of field O stars is consistent with this idea.

For the newly detected magnetic O-type stars, HD\,153426, HD\,153919, and HD\,154643, Schilbach \& Roeser (\cite{SchilbachRoeser2008}) 
suggested that the three stars were 
ejected from the clusters Hogg\,22, NGC\,6231, and ASCC\,88, respectively.
On the other hand, none of the four magnetic Of?p stars is known to belong to a cluster or an
association. The study of the evolutionary state of HD\,108 and HD\,191612 indicates that both stars are significantly 
evolved (Martins et al.\ \cite{Martins2010}). Our kinematic study of the Of?p star HD\,148937 showed that it 
possesses a space velocity of 32\,km\,s$^{-1}$ with respect to the Galactic
open cluster system, with the velocity component 
$U$=$-$26 directed opposite from the Galactic center and the velocity component $W$=$-$13
directed from the Galactic plane (Hubrig et al.\ \cite{Hubrig2011b}). 
These rather large velocities indicate that this star can be considered as a 
candidate runaway star. 

It is striking that the major part of previously detected magnetic O-type stars are 
candidate runaway stars (Hubrig et al.\ \cite{Hubrig2011b, Hubrig2011c}).  Also in the sample of O-stars with magnetic 
fields detected in this work, 
four other stars, HD\,130298, HD\,153426, HD\,153919, and HD\,154643, are mentioned in the literature as candidate 
runaway stars.  
In the past, two mechanisms  were discussed to explain the existence of runaway stars: In one 
scenario, 
close multibody interactions in a dense cluster environment cause one or more stars to be scattered 
out of the region (e.g.\ Leonard \& Duncan \cite{LeonardDuncan1990}). For this  mechanism, runaways are 
ejected in dynamical three- or four-body interactions. An alternative mechanism involves a supernova explosion 
within a close binary, ejecting the secondary due to the conservation of momentum 
(Zwicky \cite{Zwicky1957}; Blaauw \cite{Blaauw1961}). However, none of these scenarios consider the possibility how 
a massive star can acquire a magnetic field during the ejection process. 
Clearly, these findings generate a strong motivation
to carry out a kinematic study of all stars previously surveyed for magnetic fields
to search for a correlation between the kinematic status and the presence of a magnetic field.

Based on the still very limited magnetic surveys in massive stars, we cannot yet answer the question
if O-type stars are magnetic in certain evolutionary states and in a specific environment. 
Open star clusters and associations are very useful laboratories to test star formation and stellar 
evolution. The ages of our subsample of three stars with magnetic fields do not
contradict the idea, that it is drawn from the general distribution of cluster ages. 
We have to keep in mind though that 
we have a very small number statistics.
It appears that our observations
are consistent with the assumption that the presence of a magnetic field can be expected in stars of different 
classification categories. Although it was possible to recognize a few hot Of?p magnetic stars as being peculiar 
on the basis of their spectral morphology, prior to their field detection (Walborn \cite{walborn06}), 
the presence of a magnetic field can also be expected in stars of other classifications. 
Future magnetic field measurements are urgently needed to constrain the conditions controlling the 
presence of magnetic fields in hot stars, and the implications of these fields on their 
physical parameters and evolution.

{
\acknowledgements
NVK and AEP thank for support by DFG grant RO 528/10-1.
AEP acknowledges support of the RFBR grant 10-02-91338.

}

\end{document}